\pdfoutput=1
\documentclass[reprint,aps,prl,twocolumn,superscriptaddress,amsmath,amssymb,floatfix]{revtex4-2}

\usepackage{graphicx}
\usepackage{bm}
\usepackage{booktabs}
\usepackage{mathtools}
\usepackage{xcolor}
\usepackage[colorlinks=true,linkcolor=blue!65!black,citecolor=blue!65!black,urlcolor=blue!65!black]{hyperref}
\newcommand{\AmDtwoN}{7}
\newcommand{\AmDfourN}{4}
\newcommand{\AmDfiveN}{10}
\newcommand{\AmDsixN}{5}
\newcommand{\AmDsixLat}{1.07--5.27}
\newcommand{\AmEventsAll}{36}
\newcommand{\AmCtwoRangeAll}{0.482--0.522}
\newcommand{\AmLatDoneDfive}{0.37--3.63}
\newcommand{\AmTotLat}{48}
\newcommand{\AmTotDays}{90}

\newif\ifshowlogos
\showlogostrue
\ifshowlogos
  \AddToHookNext{shipout/foreground}{\put(0,0){%
    \hspace*{\dimexpr1in+\hoffset+\oddsidemargin\relax}%
    \raisebox{-22pt}[0pt][0pt]{\makebox[\textwidth]{%
      \includegraphics[height=13pt]{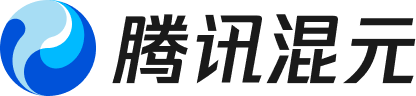}\hspace{10pt}%
      \includegraphics[height=13pt]{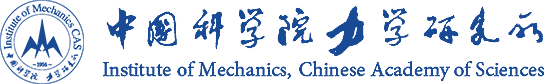}\hfill
      \includegraphics[height=13pt]{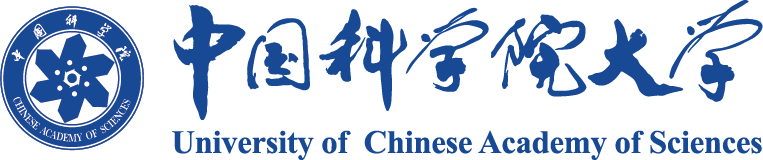}}}}}
\fi

\begin{document}

\title{Physics-Guided Recursive Self-Improvement of\\ Amortized Bayesian Inference Across Physics}

\author{Bo Liang}
\affiliation{Center for Gravitational Wave Experiment, National Microgravity Laboratory, Institute of Mechanics, Chinese Academy of Sciences, Beijing 100190, China}
\affiliation{Taiji Laboratory for Gravitational Wave Universe (Beijing/Hangzhou), University of Chinese Academy of Sciences, Beijing 100049, China}
\affiliation{Tencent Hunyuan}
\author{Mingtao Chen}
\affiliation{Tencent Hunyuan}
\author{Minghui Du}
\affiliation{Center for Gravitational Wave Experiment, National Microgravity Laboratory, Institute of Mechanics, Chinese Academy of Sciences, Beijing 100190, China}
\affiliation{Taiji Laboratory for Gravitational Wave Universe (Beijing/Hangzhou), University of Chinese Academy of Sciences, Beijing 100049, China}
\author{Ziren Luo}
\affiliation{Center for Gravitational Wave Experiment, National Microgravity Laboratory, Institute of Mechanics, Chinese Academy of Sciences, Beijing 100190, China}
\affiliation{Taiji Laboratory for Gravitational Wave Universe (Beijing/Hangzhou), University of Chinese Academy of Sciences, Beijing 100049, China}
\affiliation{Key Laboratory of Gravitational Wave Precision Measurement of Zhejiang Province, Hangzhou Institute for Advanced Study, University of Chinese Academy of Sciences, Hangzhou 310024, China}
\author{Guohua Wang}
\email[Corresponding author: ]{guohuawang@tencent.com}
\affiliation{Tencent Hunyuan}
\author{Peng Xu}
\email[Corresponding author: ]{xupeng@imech.ac.cn}
\affiliation{Center for Gravitational Wave Experiment, National Microgravity Laboratory, Institute of Mechanics, Chinese Academy of Sciences, Beijing 100190, China}
\affiliation{Taiji Laboratory for Gravitational Wave Universe (Beijing/Hangzhou), University of Chinese Academy of Sciences, Beijing 100049, China}
\affiliation{Key Laboratory of Gravitational Wave Precision Measurement of Zhejiang Province, Hangzhou Institute for Advanced Study, University of Chinese Academy of Sciences, Hangzhou 310024, China}
\affiliation{Lanzhou Center of Theoretical Physics, Lanzhou University, Lanzhou 730000, China}
\author{Zhichao Hu}
\affiliation{Tencent Hunyuan}
\author{Maxm Pan}
\affiliation{Tencent Hunyuan}

\date{\today}

\begin{abstract}
Across physics, from the orbits of imaged exoplanets and the matter inside neutron stars to thin films at synchrotrons and cortical neurons, conservation laws, interference, and compensating currents compress the posteriors of precise data into tiny, curved regions where Markov chains need minutes to days. We introduce \textsc{PhyRSI} (physics-guided recursive self-improvement), in which an AI agent coupled to the simulator writes and revises an amortized Bayesian inference algorithm for each field, scored by the importance weights of the exact likelihood because the true posterior is unknown. It evolves coordinates, symmetries, and factorizations matched to the physics: orbital elements at the best-measured epoch, Fresnel-normalized reflectivity with a chart for each of its multiple solutions, coordinates along which Na$^{+}$ and K$^{+}$ currents compensate, a shared neutron-star equation of state that factorizes a $32$-dimensional joint analysis, and the time-rescaling symmetry of predator--prey dynamics. In six fields, networks trained once on simulations return \AmEventsAll\ posteriors of real observations in seconds, reweighted by the likelihood and agreeing with converged Markov chains about as closely as two independent chains agree with each other. Such posteriors could steer synchrotron film growth and patch-clamp recordings as they run, follow X-ray transients while they are bright, update the dense-matter equation of state as neutron stars are added, and compare physical models across catalogs. Inference can thus be evolved and checked by the physics itself rather than designed by hand for each problem.
\end{abstract}

\raggedbottom
\maketitle

\onecolumngrid
\begin{center}\begin{minipage}{\textwidth}\makeatletter\def\@captype{figure}\makeatother
\centering
\includegraphics[width=0.87\textwidth]{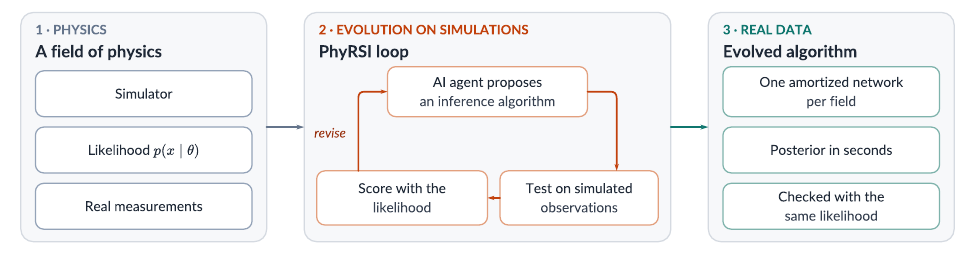}
\caption{\textsc{PhyRSI}. Coupled to the simulator and likelihood of a field (left), an AI agent proposes an inference algorithm, scores it with the likelihood weights, and revises it until it passes on held-out simulations (center); the evolved algorithm returns the posterior of each real observation in seconds, checked with the same likelihood (right).}
\label{fig:workflow}
\end{minipage}\end{center}
\twocolumngrid

\textit{Introduction.}---Physics turns data into knowledge by solving inverse problems. Physical laws map parameters forward to signals, so the equation of state of supranuclear matter, the flow of gas onto a compact object, the structure of a buried interface, and the conductances of an excitable membrane are never read off a detector but recovered by inverting a forward model into a posterior distribution over the parameters~\cite{Jaynes1957,Tarantola2005,Cranmer2020}. Observatories, light sources, and automated laboratories now deliver data that are both precise and plentiful~\cite{Abbott2017PRL,Ivezic2019,Noack2021,Szymanski2023}, and a single measurement routinely carries tens of nats of information about its parameters~\cite{Lindley1956,Braunstein1994}. Increasingly the inversion is needed while phenomena unfold and across whole populations, to characterize transients and mergers as they happen~\cite{Abbott2017ApJL,Coulter2017,Fabian2009,Kara2019}, to map the dense-matter equation of state with growing samples of neutron stars~\cite{Lattimer2004,Abbott2018PRL,Annala2018,Annala2020,Dietrich2020,Huth2022}, and to steer materials synthesis with feedback from the measurement~\cite{Parratt1954,Burger2020,Noack2021,Szymanski2023}. %
\begin{table*}[t!]
\centering
\caption{Inference algorithms evolved by \textsc{PhyRSI} in six fields and the potential impact of second-scale inference; accuracy and timing in Table~\ref{tab:jsd}.}
\label{tab:physical_significance}
\renewcommand{\arraystretch}{1.18}
\setlength{\tabcolsep}{4.5pt}
\footnotesize
\begin{tabular}{@{}p{0.165\textwidth}p{0.19\textwidth}p{0.235\textwidth}p{0.357\textwidth}@{}}
\toprule
\textbf{Field and data} & \textbf{Physical question} & \textbf{Algorithm \textsc{PhyRSI} evolved} & \textbf{Potential impact of second-scale inference} \\
\midrule
\raggedright \textbf{D1 Celestial mechanics} \newline Orbits of three imaged planets ($8\mathrm{D}$)~\cite{Lagrange2010,Snellen2014,Nielsen2020,Blunt2017,Blunt2020} &
\raggedright Three-dimensional orbit and total mass from relative astrometry ($7$--$34$ epochs over $1$--$15$~yr). &
\raggedright Equinoctial elements and mean longitude at the best-measured epoch (for $\beta$~Pic~b, the orbital radius at that epoch in place of $a$); one network per system. &
\raggedright Re-analysis in seconds makes routine the thousands of injection--recovery fits that debias short-arc eccentricities~\cite{FerrerChavez2021}, which could help tell whether wide giant planets formed in disks or by cloud fragmentation~\cite{Bowler2020}. \tabularnewline
\addlinespace[2pt]
\raggedright \textbf{D2 X-ray astrophysics} \newline \emph{XMM-Newton} and \emph{Chandra} spectra (\AmDtwoN~sources,~$5\mathrm{D}$)~\cite{Fabian2009,Kara2019,Arnaud1996,Barret2024} &
\raggedright Absorption column and continuum (power law plus blackbody), folded through each detector response. &
\raggedright Response-conditioned network in log-flux coordinates; follows the degeneracy between absorption and the soft continuum ($N_H$, $\Gamma$, $kT_{\mathrm{bb}}$). &
\raggedright Posteriors within seconds could separate thermal from nonthermal emission in \emph{Einstein Probe} transients~\cite{Yuan2025} while they are bright, flagging supernova shock breakouts~\cite{Soderberg2008} and tidal disruptions~\cite{Gezari2021} for follow-up. \tabularnewline
\addlinespace[2pt]
\raggedright \textbf{D3 Dense matter} \newline Six neutron stars in globular clusters ($7\mathrm{D}$;~jointly~$32\mathrm{D}$)~\cite{Heinke2006,Guillot2013,Steiner2018,Brandes2024} &
\raggedright Equation of state of supranuclear matter from hydrogen-atmosphere spectra of quiescent neutron stars, star by star and jointly. &
\raggedright Factorization into a shared equation of state and per-star parameters, which removes the collapse of joint importance weights. &
\raggedright Adding stars updates the shared equation of state in seconds, for combination with \emph{NICER}~\cite{Riley2019} and gravitational-wave data~\cite{Abbott2018PRL}; a kink in the mass--radius relation could signal a first-order transition to quark matter~\cite{Alford2013,Annala2020}. \tabularnewline
\addlinespace[2pt]
\raggedright \textbf{D4 Condensed matter} \newline X-ray reflectivity of thin films (\AmDfourN~curves,~$6\mathrm{D}$)~\cite{Parratt1954,Majkrzak1995,Nelson2019,Starostin2024} &
\raggedright Film thickness, interface roughness, and scattering-length densities from specular reflectivity (Parratt recursion). &
\raggedright Fresnel-normalized curve ($Rq^4$) and local Laplace charts with adaptive importance sampling for the multiple solutions left by Kiessig fringes. &
\raggedright Posteriors after each few-second scan at synchrotron beamlines could measure the roughening of a growing film \emph{in situ}, testing Kardar--Parisi--Zhang scaling~\cite{Kardar1986}, and steer autonomous deposition toward smoother films~\cite{Noack2021}. \tabularnewline
\addlinespace[2pt]
\raggedright \textbf{D5 Biophysics} \newline Hodgkin--Huxley neurons, Allen Cell Types (\AmDfiveN~cells,~$7\mathrm{D}$)~\cite{HodgkinHuxley1952,Pospischil2008,Prinz2004,Goncalves2020} &
\raggedright Na$^{+}$, K$^{+}$, M-type, and leak conductances, gating time constant, spike threshold, and leak reversal potential from ten voltage features of current-clamp recordings. &
\raggedright Network conditioned on the stimulus protocol, with a learned affine chart in logit-conductance coordinates; captures the compensation between Na$^{+}$ and K$^{+}$ currents. &
\raggedright Conductance posteriors during the recording could let experimenters choose the next stimulus, a channel blocker, or a dynamic-clamp conductance~\cite{Prinz2004TINS} to separate compensating Na$^{+}$ and K$^{+}$ currents while the cell is still patched~\cite{MarderGoaillard2006}. \tabularnewline
\addlinespace[2pt]
\raggedright \textbf{D6 Population dynamics} \newline Predator--prey time series (\AmDsixN~series,~$8\mathrm{D}$)~\cite{Lotka1920,Volterra1926,Lueckmann2021} &
\raggedright Interaction rates, initial populations, and noise of the Lotka--Volterra equations from hare--lynx, mite, protozoan, and wolf--moose counts. &
\raggedright Coordinates aligned with the exact time-rescaling symmetry, with the mixture network shifted to each aliased period; captures period aliasing and mass piled against the prior boundary. &
\raggedright Marginal likelihoods from the importance weights could compare functional responses~\cite{Holling1959} across ecological records, testing whether population cycles come from predator--prey coupling or from single-species feedback~\cite{Murdoch2002}. \tabularnewline
\bottomrule
\end{tabular}
\end{table*}%
\begin{table*}[t!]
\centering
\caption{Accuracy and speed of \textsc{PhyRSI} on real observations. Tiles: one-dimensional marginal Jensen--Shannon divergences (mnat; logarithmic shading, darker: larger) from a converged MCMC reference; \textsc{PhyRSI}--MCMC is their mean over the parameters of an event (for D3 joint, the stellar columns show the maximum over the six stars), and MCMC--MCMC the same mean between two references. C2ST: classifier two-sample test ($0.5$: indistinguishable). $t$: \textsc{PhyRSI} time per event on H20 GPUs; 4-core MCMC: one reference chain on four CPU cores; speedup: their ratio. Orbital coordinates $h$, $k$, $\lambda_{\mathrm{ref}}$: End Matter.}
\label{tab:jsd}
\providecommand{\jt}[2]{{\setlength{\fboxsep}{0.9pt}\colorbox{blue!#2}{\makebox[2.85em]{#1}}}}
\providecommand{\domrow}[2]{\multicolumn{15}{@{}l@{}}{\makebox[0pt][l]{\textbf{#1}\quad #2}}}
\scriptsize
\setlength{\tabcolsep}{0pt}
\renewcommand{\arraystretch}{0.92}
\begin{tabular*}{\textwidth}{@{}l@{\extracolsep{\fill}}c*{7}{@{\extracolsep{0pt}\hspace{3.2pt}}c}@{\extracolsep{\fill}}c@{\extracolsep{0pt}\hspace{8pt}}c@{\extracolsep{\fill}}crrr@{}}
\toprule
\multicolumn{9}{@{}l}{\makebox[0pt][l]{\textbf{D1 Exoplanet orbits ($8\mathrm{D}$)}\quad 1 GPU; one network per planetary system}} & \multicolumn{2}{c}{mean JSD (mnat)} & & & & \\
\cmidrule(lr){10-11}
 & & & & & & & & & \textsc{PhyRSI}-- & MCMC-- & & & 4-core & \\
 & $a$ & $e$ & $i$ & $\omega$ & $\Omega$ & $\tau$ & $\varpi$ & $M_{\mathrm{tot}}$ & MCMC & MCMC & C2ST & $t$ (s) & MCMC & speedup \\[1pt]
$\beta$~Pic~b & \jt{0.17}{15} & \jt{0.13}{11} & \jt{0.19}{16} & \jt{0.55}{29} & \jt{0.13}{12} & \jt{0.41}{25} & \jt{0.24}{19} & \jt{0.14}{12} & \jt{0.25}{19} & \jt{0.22}{18} & 0.492 & 0.48 & 26.2\,min & 3{,}280$\times$ \\
 & $a$ & $h$ & $k$ & $\cos i$ & $\Omega$ & $\lambda_{\mathrm{ref}}$ & $\varpi$ & $M_{\mathrm{tot}}$ & & & & & & \\[1pt]
GJ~504~b & \jt{1.46}{40} & \jt{1.02}{36} & \jt{0.23}{18} & \jt{0.30}{21} & \jt{0.71}{32} & \jt{0.91}{35} & \jt{0.40}{25} & \jt{0.28}{20} & \jt{0.66}{31} & \jt{0.77}{33} & 0.522 & 0.68 & 7.3\,min & 648$\times$ \\
HD~206893~B & \jt{0.94}{35} & \jt{0.65}{31} & \jt{0.08}{5} & \jt{0.29}{21} & \jt{0.16}{14} & \jt{0.12}{11} & \jt{0.24}{19} & \jt{0.20}{16} & \jt{0.34}{23} & \jt{0.55}{29} & 0.499 & 3.63 & 7.7\,min & 127$\times$ \\
\midrule
\domrow{D2 X-ray spectra ($5\mathrm{D}$)}{1 GPU} \\
 & $N_H$ & $\Gamma$ & $\log K_{\mathrm{pl}}$ & $\log kT_{\mathrm{bb}}$ & $\log K_{\mathrm{bb}}$ &  &  &  & & & & & & \\[1pt]
ulx4\_xmm & \jt{0.21}{17} & \jt{0.13}{11} & \jt{0.25}{19} & \jt{0.18}{15} & \jt{0.21}{17} &  &  &  & \jt{0.19}{16} & \jt{0.18}{15} & 0.501 & 0.74 & 4.93\,h & 24{,}000$\times$ \\
csc\_47Tuc\_X7 & \jt{0.26}{20} & \jt{0.33}{23} & \jt{0.29}{21} & \jt{0.32}{22} & \jt{0.15}{13} &  &  &  & \jt{0.27}{20} & \jt{0.19}{16} & 0.501 & 1.20 & 1.65\,h & 4{,}940$\times$ \\
csc\_NGC6397\_U24 & \jt{0.15}{13} & \jt{0.24}{19} & \jt{0.22}{17} & \jt{0.14}{12} & \jt{0.20}{16} &  &  &  & \jt{0.19}{16} & \jt{0.18}{15} & 0.502 & 0.61 & 1.58\,h & 9{,}360$\times$ \\
csc\_M30\_A1 & \jt{0.24}{19} & \jt{0.18}{16} & \jt{0.25}{19} & \jt{0.36}{24} & \jt{0.24}{19} &  &  &  & \jt{0.26}{20} & \jt{0.16}{14} & 0.509 & 0.61 & 1.57\,h & 9{,}230$\times$ \\
csc\_NGC6304\_X4 & \jt{0.10}{9} & \jt{0.07}{5} & \jt{0.23}{18} & \jt{0.14}{12} & \jt{0.16}{14} &  &  &  & \jt{0.14}{12} & \jt{0.13}{12} & 0.493 & 0.61 & 1.62\,h & 9{,}580$\times$ \\
csc\_M13 & \jt{0.15}{13} & \jt{0.14}{13} & \jt{0.15}{13} & \jt{0.18}{16} & \jt{0.17}{15} &  &  &  & \jt{0.16}{14} & \jt{0.12}{11} & 0.515 & 0.61 & 1.54\,h & 9{,}150$\times$ \\
csc\_omega\_Cen & \jt{0.17}{15} & \jt{0.12}{10} & \jt{0.21}{17} & \jt{0.13}{12} & \jt{0.28}{21} &  &  &  & \jt{0.18}{15} & \jt{0.13}{11} & 0.505 & 1.20 & 1.56\,h & 4{,}660$\times$ \\
\midrule
\domrow{D3 Neutron stars, single ($7\mathrm{D}$)}{1 GPU; joint analysis ($32\mathrm{D}$) on 2 GPUs} \\
 & $\lambda_1$ & $\lambda_2$ & $N_H$ & $d$ & $\log T$ & $u$ & $K$ &  & & & & & & \\[1pt]
X7 & \jt{0.29}{21} & \jt{0.19}{16} & \jt{0.30}{22} & \jt{0.36}{24} & \jt{0.20}{16} & \jt{0.20}{17} & \jt{0.18}{16} &  & \jt{0.25}{19} & \jt{0.17}{15} & 0.510 & 0.52 & 38.8\,min & 4{,}510$\times$ \\
U24 & \jt{0.20}{16} & \jt{0.15}{13} & \jt{0.19}{16} & \jt{0.42}{25} & \jt{0.20}{16} & \jt{0.14}{12} & \jt{0.16}{14} &  & \jt{0.21}{17} & \jt{0.18}{15} & 0.497 & 0.46 & 25.5\,min & 3{,}330$\times$ \\
M30 & \jt{0.29}{21} & \jt{0.11}{9} & \jt{0.08}{6} & \jt{0.13}{12} & \jt{0.13}{12} & \jt{0.14}{12} & \jt{0.19}{16} &  & \jt{0.15}{13} & \jt{0.15}{13} & 0.498 & 0.53 & 46.1\,min & 5{,}230$\times$ \\
N6304 & \jt{0.11}{10} & \jt{0.24}{19} & \jt{0.22}{17} & \jt{0.40}{25} & \jt{0.10}{8} & \jt{0.13}{11} & \jt{0.24}{19} &  & \jt{0.21}{17} & \jt{0.17}{14} & 0.500 & 0.39 & 15.5\,min & 2{,}410$\times$ \\
M13 & \jt{0.21}{17} & \jt{0.10}{8} & \jt{0.11}{9} & \jt{0.22}{17} & \jt{0.13}{12} & \jt{0.05}{1} & \jt{0.34}{23} &  & \jt{0.17}{14} & \jt{0.13}{11} & 0.501 & 0.73 & 15.7\,min & 1{,}280$\times$ \\
wCen & \jt{0.08}{6} & \jt{0.14}{12} & \jt{0.33}{22} & \jt{0.21}{17} & \jt{0.12}{10} & \jt{0.07}{4} & \jt{0.30}{21} &  & \jt{0.18}{15} & \jt{0.13}{11} & 0.501 & 0.37 & 18.2\,min & 2{,}910$\times$ \\
D3 joint & \jt{0.21}{17} & \jt{0.52}{28} & \jt{0.30}{22} & \jt{0.28}{20} & \jt{0.26}{20} & \jt{0.24}{19} & \jt{0.33}{23} &  & \jt{0.21}{17} & \jt{0.21}{17} & 0.507 & 1.66 & 2.66\,h & 5{,}780$\times$ \\
\midrule
\domrow{D4 X-ray reflectivity ($6\mathrm{D}$)}{1 GPU} \\
 & $d_{\mathrm{film}}$ & $\sigma_{\mathrm{top}}$ & $\sigma_{\mathrm{sub}}$ & $\rho_{\mathrm{film}}$ & $\rho_{\mathrm{sub}}$ & $\log b$ &  &  & & & & & & \\[1pt]
C60 & \jt{0.20}{17} & \jt{0.12}{10} & \jt{0.12}{10} & \jt{0.15}{13} & \jt{0.25}{19} & \jt{0.19}{16} &  &  & \jt{0.17}{15} & \jt{0.15}{13} & 0.498 & 0.57 & 15.1\,min & 1{,}580$\times$ \\
PTCDI\_C3 & \jt{0.23}{18} & \jt{0.22}{17} & \jt{0.22}{17} & \jt{0.17}{14} & \jt{0.13}{12} & \jt{0.30}{22} &  &  & \jt{0.21}{17} & \jt{0.21}{17} & 0.517 & 1.13 & 24.3\,min & 1{,}300$\times$ \\
DIP\_34a & \jt{0.19}{16} & \jt{0.14}{12} & \jt{0.18}{15} & \jt{0.20}{17} & \jt{0.26}{20} & \jt{0.12}{10} &  &  & \jt{0.18}{15} & \jt{0.14}{13} & 0.504 & 1.15 & 40.2\,min & 2{,}100$\times$ \\
S33a & \jt{0.27}{20} & \jt{0.22}{18} & \jt{0.19}{16} & \jt{0.28}{20} & \jt{0.16}{14} & \jt{0.20}{16} &  &  & \jt{0.22}{18} & \jt{0.19}{16} & 0.500 & 1.73 & 44.2\,min & 1{,}540$\times$ \\
\midrule
\domrow{D5 Hodgkin--Huxley neurons ($7\mathrm{D}$)}{1 GPU; MCMC with the NumPy simulator} \\
 & $\bar g_{\mathrm{Na}}$ & $\bar g_{\mathrm{K}}$ & $g_{\mathrm{L}}$ & $\bar g_{\mathrm{M}}$ & $\tau_{\max}$ & $V_T$ & $E_{\mathrm{L}}$ &  & & & & & & \\[1pt]
c518290966 & \jt{0.30}{21} & \jt{0.22}{18} & \jt{0.18}{15} & \jt{0.29}{21} & \jt{0.07}{5} & \jt{0.20}{16} & \jt{0.54}{28} &  & \jt{0.26}{20} & \jt{0.23}{18} & 0.489 & 0.93 & 206\,h & 794{,}000$\times$ \\
c509881736 & \jt{0.21}{17} & \jt{0.18}{15} & \jt{0.21}{17} & \jt{0.18}{15} & \jt{0.11}{10} & \jt{0.11}{9} & \jt{0.17}{15} &  & \jt{0.17}{14} & \jt{0.11}{10} & 0.497 & 0.84 & 206\,h & 889{,}000$\times$ \\
c566517779 & \jt{0.20}{16} & \jt{0.21}{17} & \jt{0.25}{19} & \jt{0.16}{14} & \jt{0.14}{12} & \jt{0.17}{15} & \jt{0.15}{13} &  & \jt{0.18}{15} & \jt{0.19}{16} & 0.482 & 0.82 & 205\,h & 902{,}000$\times$ \\
c567399060 & \jt{0.29}{21} & \jt{0.30}{22} & \jt{0.10}{8} & \jt{0.11}{9} & \jt{0.06}{1} & \jt{0.09}{7} & \jt{0.46}{27} &  & \jt{0.20}{17} & \jt{0.17}{15} & 0.508 & 0.82 & 206\,h & 906{,}000$\times$ \\
c569469018 & \jt{0.13}{12} & \jt{0.22}{17} & \jt{0.10}{9} & \jt{0.16}{14} & \jt{0.10}{8} & \jt{0.14}{12} & \jt{0.06}{3} &  & \jt{0.13}{11} & \jt{0.16}{14} & 0.504 & 0.82 & 205\,h & 897{,}000$\times$ \\
c532571720 & \jt{0.18}{15} & \jt{0.16}{14} & \jt{0.24}{19} & \jt{0.30}{21} & \jt{0.08}{5} & \jt{0.29}{21} & \jt{0.19}{16} &  & \jt{0.21}{17} & \jt{0.16}{14} & 0.495 & 0.82 & 205\,h & 900{,}000$\times$ \\
c555060623 & \jt{0.25}{19} & \jt{0.39}{25} & \jt{0.24}{19} & \jt{0.33}{23} & \jt{0.13}{12} & \jt{0.22}{18} & \jt{0.11}{10} &  & \jt{0.24}{19} & \jt{0.15}{13} & 0.500 & 0.83 & 207\,h & 903{,}000$\times$ \\
c534524026 & \jt{0.20}{17} & \jt{0.13}{12} & \jt{0.32}{22} & \jt{0.20}{16} & \jt{0.10}{8} & \jt{0.23}{18} & \jt{0.11}{9} &  & \jt{0.18}{16} & \jt{0.14}{12} & 0.507 & 0.82 & 205\,h & 896{,}000$\times$ \\
c532355382 & \jt{0.23}{18} & \jt{0.18}{15} & \jt{0.18}{16} & \jt{0.19}{16} & \jt{0.10}{8} & \jt{0.17}{15} & \jt{0.14}{12} &  & \jt{0.17}{15} & \jt{0.13}{11} & 0.493 & 0.84 & 206\,h & 879{,}000$\times$ \\
c526950199 & \jt{0.13}{11} & \jt{0.13}{11} & \jt{0.20}{17} & \jt{0.24}{19} & \jt{0.10}{8} & \jt{0.20}{17} & \jt{0.07}{3} &  & \jt{0.15}{13} & \jt{0.16}{14} & 0.505 & 0.82 & 208\,h & 910{,}000$\times$ \\
\midrule
\domrow{D6 Lotka--Volterra series ($8\mathrm{D}$)}{8 shared GPUs} \\
 & $\ln\alpha$ & $\ln\beta$ & $\ln\gamma$ & $\ln\delta$ & $\ln u_0$ & $\ln v_0$ & $\ln\sigma_u$ & $\ln\sigma_v$ & & & & & & \\[1pt]
hare\_lynx & \jt{0.26}{20} & \jt{0.24}{19} & \jt{0.18}{15} & \jt{0.13}{12} & \jt{0.16}{14} & \jt{0.19}{16} & \jt{0.17}{15} & \jt{0.17}{14} & \jt{0.19}{16} & \jt{0.15}{13} & 0.492 & 4.40 & 7.33\,h & 6{,}000$\times$ \\
huffaker\_60w & \jt{0.29}{21} & \jt{0.23}{18} & \jt{0.24}{19} & \jt{0.17}{15} & \jt{0.24}{19} & \jt{0.39}{25} & \jt{0.12}{10} & \jt{0.21}{17} & \jt{0.24}{19} & \jt{0.20}{17} & 0.498 & 4.64 & 21.6\,h & 16{,}700$\times$ \\
gause\_f39 & \jt{0.32}{22} & \jt{0.48}{27} & \jt{0.19}{16} & \jt{0.26}{20} & \jt{0.18}{15} & \jt{0.24}{19} & \jt{0.13}{12} & \jt{0.16}{14} & \jt{0.25}{19} & \jt{0.20}{16} & 0.506 & 4.27 & 6.18\,h & 5{,}210$\times$ \\
isle\_royale & \jt{1.04}{36} & \jt{0.98}{36} & \jt{0.96}{35} & \jt{0.66}{31} & \jt{0.22}{18} & \jt{0.21}{17} & \jt{0.34}{23} & \jt{0.27}{20} & \jt{0.59}{29} & \jt{0.31}{22} & 0.517 & 1.07 & 25.8\,h & 86{,}600$\times$ \\
gause\_science & \jt{0.06}{2} & \jt{0.09}{6} & \jt{0.14}{12} & \jt{0.09}{7} & \jt{0.23}{18} & \jt{0.46}{27} & \jt{0.22}{18} & \jt{0.22}{18} & \jt{0.19}{16} & \jt{0.15}{13} & 0.489 & 5.27 & 5.85\,h & 4{,}000$\times$ \\
\bottomrule
\end{tabular*}
\end{table*}
\begin{figure*}[t!]
\centering
\includegraphics[width=0.80\textwidth]{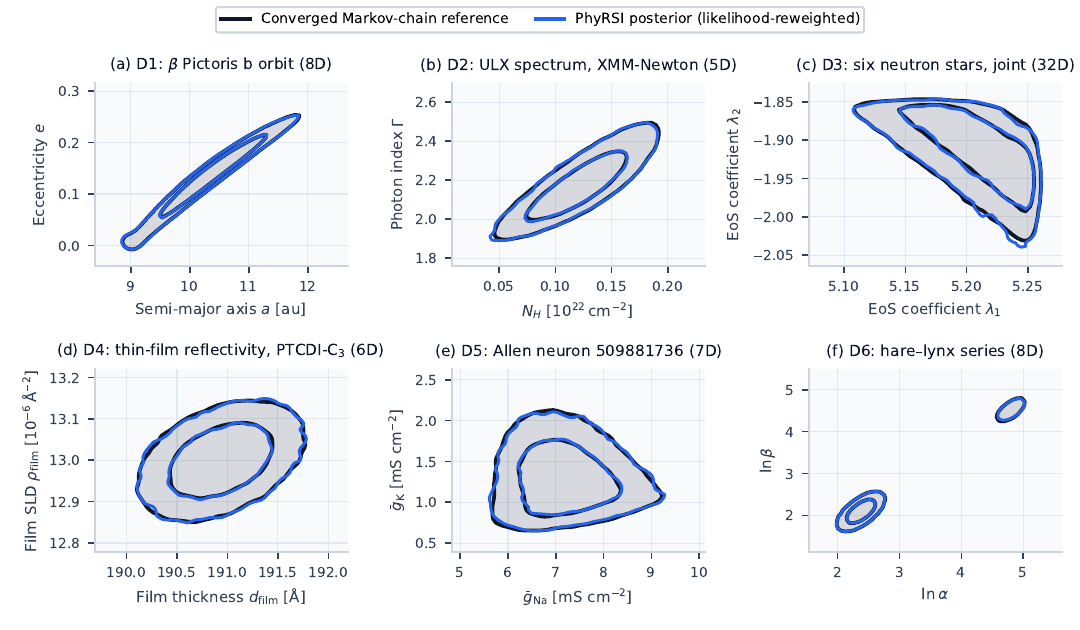}
\caption{Posteriors of real observations in six fields: $68\%$ and $95\%$ credible contours of two-dimensional marginals from a converged Markov-chain reference (dark, shaded) and from \textsc{PhyRSI} after likelihood reweighting (blue). In (f), rates are in units of the inverse series duration, and two aliased oscillation periods give two modes. Joint posteriors of all \AmEventsAll\ observations are in the Supplemental Material.}
\label{fig:six_domains}
\end{figure*}%
Across physics, the bottleneck is shifting from acquiring data to inverting them.

Precision makes the inversion hard. Conservation laws, interference, and compensating currents leave a few stiff parameter combinations and many sloppy ones~\cite{Majkrzak1995,Prinz2004,MarderGoaillard2006,Waterfall2006,Transtrum2010,Machta2013}, so the posterior fills a fraction $V_{\mathrm{post}}/V_{\mathrm{prior}}\sim e^{-D_{\mathrm{KL}}}$ of the prior volume---$4\times10^{-11}$ for the orbit of $\beta$~Pictoris~b---with curvatures that differ by six to ten orders of magnitude. Markov chains need $10^{7}$--$10^{8}$ sequential likelihood evaluations to cross such narrow, curved valleys~\cite{Metropolis1953,Hukushima1996,Duane1987,ForemanMackey2013}, while the prior simulations used to train neural estimators~\cite{Papamakarios2016,Durkan2019,Lipman2023} almost never land inside the posterior, a rare-event problem akin to estimating free energies from nonequilibrium work~\cite{Jarzynski1997,Kawai2007,Chatterjee2018}. Generic neural posteriors can be severely overconfident, and calibration averaged over the prior certifies no single observation~\cite{Hermans2022,Talts2018}. Where fast and reliable inference exists, it rests on years of field-specific design: the symmetries of gravitational-wave signals~\cite{Dax2021PRL,Dax2023PRL,Gabbard2022,Dax2025Nature}, internal coordinates of molecules~\cite{Noe2019}, and gauge equivariance in lattice field theory~\cite{Wu2019,Kanwar2020}. Each new system has required its own insight.

Here we introduce \textsc{PhyRSI} (\emph{Physics-guided Recursive Self-Improvement}~\cite{Good1965}), in which an AI agent autonomously evolves the amortized inference algorithm of a field, scored by the physical likelihood (Fig.~\ref{fig:workflow}). Computational physics has repeatedly defeated sampling barriers by finding the right variables, from cluster moves at criticality~\cite{SwendsenWang1987} to conjugate momenta in lattice gauge theory~\cite{Duane1987}, and self-improving search has surpassed human design wherever candidates can be scored~\cite{Silver2017,Real2020,Fawzi2022,Mankowitz2023,RomeraParedes2024,Novikov2025}. In inference the true posterior is unknown, but the physical likelihood supplies the score: the posterior $p(\bm{\theta}\mid\bm{x})$ is a Boltzmann distribution with energy $-\ln[p(\bm{\theta})\,p(\bm{x}\mid\bm{\theta})]$, and for $M$ draws from any candidate $q(\bm{\theta}\mid\bm{x})$ the importance weights $w\propto p(\bm{\theta})\,p(\bm{x}\mid\bm{\theta})/q(\bm{\theta}\mid\bm{x})$ measure the overlap of the two ensembles, as in free-energy perturbation~\cite{Jarzynski1997,Zwanzig1954},
\begin{equation}
D_{\mathrm{KL}}(p\,\|\,q)\le\ln\!\left[1+\chi^{2}(p\,\|\,q)\right]\approx-\ln(\mathrm{ESS}/M),
\label{eq:verifier}
\end{equation}
where ESS is the effective sample size, and the tail index $\hat{k}$ of the largest weights flags posterior mass that $q$ underweights~\cite{Kong1992,Agapiou2017,Vehtari2024}. Because the estimate comes from the draws themselves, it cannot see a mode that no draw reaches, so passing it is necessary, not sufficient, for a correct posterior. Coupled to the simulator and likelihood, the agent writes an inference algorithm, tests it on simulated observations, reads from the weights why it fails, and rewrites its coordinates, symmetries, and factorization until it passes the weight check and a calibration test with known parameters on held-out simulations. The result is an amortized algorithm for that field in which a network trained once on simulations (one per planetary system for orbits), combined where needed with a local coordinate chart fitted to the observation, proposes the posterior of any new observation and the exact likelihood reweights and checks it.

\textit{What \textsc{PhyRSI} evolves.}---What the loop evolves mirrors the physics it is coupled to (Table~\ref{tab:physical_significance}; End Matter). It found coordinates matched to each field, such as equinoctial elements at the best-measured epoch (D1), the reflectivity multiplied by $q^{4}$ to remove its Fresnel decay (D4), and directions along which Na$^{+}$ and K$^{+}$ currents compensate (D5), combined where needed with a local chart or a learned affine layer that absorbs the contraction of the posterior volume and leaves the network only the anharmonic remainder. It also found the structure of each problem: stars that are independent once their shared equation of state is fixed (D3), and the exact time-rescaling symmetry of the Lotka--Volterra flow (D6).

\textit{Real observations.}---We trained amortized networks on simulations that span each field's experimental conditions, one per planetary system in D1 and one per field in D2--D6, and applied them without retraining to all \AmEventsAll\ real observations in Table~\ref{tab:jsd} (sources in the Supplemental Material). All their posteriors pass the likelihood check. Their mean divergence from a converged Markov-chain reference, $0.13$--$0.66$~mnat, is comparable to the $0.11$--$0.77$~mnat between two references with different random seeds and within twice that value for every observation~\cite{RomeroShaw2020,Dax2021PRL}, with classifier two-sample scores of \AmCtwoRangeAll\ ($0.5$: indistinguishable; Table~\ref{tab:jsd}, Fig.~\ref{fig:six_domains}). The reweighted posteriors carry the physics of each problem: the trade-off between absorption and photon index, the compensating Na$^{+}$ and K$^{+}$ conductances, two modes from aliased population cycles (Fig.~\ref{fig:six_domains}), and, in the reflectivity curves, mass piled against a prior bound, a sign that one layer is too simple for these films (Supplemental Material). A posterior takes \AmLatDoneDfive~s on one or two NVIDIA H20 GPUs in D1--D5 and \AmDsixLat~s on eight shared GPUs in D6; together, the \AmEventsAll\ posteriors take \AmTotLat~s, against about \AmTotDays~days for one four-core Markov chain per observation (Supplemental Material).

All algorithm versions were finalized before their first run on real data and compared with the references only afterwards. We chose them by the likelihood check alone in D1--D5 (two retrained D1 networks by a rule set before their training) and, in D6, as the fastest version that passes the check on all five series (Supplemental Material).

\textit{Outlook.}---\textsc{PhyRSI} needs only a forward model and a likelihood; the six fields here have nothing else in common. The same loop applies to models with up to tens of parameters wherever the likelihood can be evaluated about $10^{5}$ times per second in parallel on accelerators, from gravitational waves and cosmology to neutrino oscillations, neutron scattering, plasma diagnostics, and quantum devices, though not to purely implicit or minute-scale simulators. Earlier self-improving searches optimized objectives that can be computed directly~\cite{RomeraParedes2024,Novikov2025,Wang2025}; to our knowledge, \textsc{PhyRSI} is the first to evolve Bayesian inference algorithms scored by the physical likelihood, where the correct answer is unknown, and to validate them on real observations in several fields. Likelihood-checked posteriors in seconds would let inference steer experiments as they run and test competing models across catalogs, and the time from a new measurement to a tested physical conclusion would be set by computing, not by years of method development.

\begin{acknowledgments}
The code of all algorithm versions, the evaluation scripts, and the details of every run are available at \url{https://github.com/aspirinliang/PhyRSI}. B.L. carried out this work during an internship at Tencent Hunyuan. This work is supported by the National Key R\&D Program of China under Grant No.~2025YFE0217300 and the International Partnership Program of the Chinese Academy of Sciences (Grant No.~025GJHZ2023106GC).
\end{acknowledgments}

\appendix
\setcounter{secnumdepth}{1}

\section{The Concentration Barrier and the Likelihood Check}\label{app:barrier}
Let $p(\bm{\theta})$ be the prior on $\Theta\subset\mathbb{R}^D$ and $p(\bm{x}\mid\bm{\theta})$ the likelihood, with Fisher information $\mathcal{I}(\bm{\theta})=\mathbf{J}^\top\bm{\Sigma}_x^{-1}\mathbf{J}$ and $\mathbf{J}=\partial\mathcal{F}/\partial\bm{\theta}$ the Jacobian of the forward model $\mathcal{F}$. Precise data create three obstructions for estimators trained on prior simulations. (i)~\textit{Volume contraction}: the effective posterior volume relative to the prior is
\begin{equation}
V_{\mathrm{post}}/V_{\mathrm{prior}}\asymp\exp\!\left[-D_{\mathrm{KL}}\!\left(p(\bm{\theta}\mid\bm{x})\,\|\,p(\bm{\theta})\right)\right],
\label{eq:volume_contraction}
\end{equation}
$3.9\times10^{-11}$ for $\beta$~Pictoris~b ($D_{\mathrm{KL}}\approx24$~nats), so prior simulations essentially never land in the credible region. (ii)~\textit{Anisotropic stiffness}: the local Hessian $\mathbf{H}=\mathcal{I}(\bm{\theta}_{\mathrm{MAP}})+\mathbf{P}_0$ has condition numbers of $10^{6}$--$10^{10}$, so a transport map from a standard normal must change the log density by $\sim D_{\mathrm{KL}}$ while its velocity gradients grow as $\sqrt{\kappa(\mathbf{H})}$. (iii)~\textit{Weight collapse in hierarchical systems}: when global parameters are shared by $N$ objects, per-object proposal mismatch $\sigma^2$ adds up, $\mathrm{Var}[\ln w]\propto N\sigma^2$, and the importance-sampling efficiency decays as $\exp(-N\sigma^2)$~\cite{Chatterjee2018}.

For $M$ draws $\bm{\theta}_m\sim q$ with self-normalized weights $\bar{w}_m\propto p(\bm{\theta}_m)p(\bm{x}\mid\bm{\theta}_m)/q(\bm{\theta}_m\mid\bm{x})$, the normalized effective sample size $\mathrm{ESS}/M=1/(M\sum_m\bar{w}_m^2)$ converges to $[1+\chi^2(p\,\|\,q)]^{-1}$~\cite{Kong1992,Agapiou2017}, and Jensen's inequality under $p$, $D_{\mathrm{KL}}(p\,\|\,q)=\mathbb{E}_p\ln(p/q)\le\ln\mathbb{E}_p(p/q)$, gives Eq.~\eqref{eq:verifier}. With the Pareto tail index $\hat{k}$ of the largest weights~\cite{Vehtari2024} and a defensive mixture that adds a small prior component to the proposal~\cite{Hesterberg1995,Owen2000}, a posterior passes the likelihood check if $\mathrm{ESS}\ge2\times10^{4}$, $\hat{k}\le0.7$, the largest normalized weight is at most $10^{-3}$, and the weight carried by the prior component is at most $10^{-3}$; in D6 a probe for local maxima missed by the proposal is added. These quantities are estimated from the draws, so a mode that no draw reaches is invisible to them; this is why every posterior was also compared with long Markov chains.

\section{Evolved Operators}\label{app:operators}
The loop of Fig.~\ref{fig:workflow} runs on freshly drawn simulated observations, and software blocks any reading of true parameters or reference chains. The agent reads the simulator, the prior, and the statistics of each failure ($\mathrm{ESS}/M$, $\hat{k}$, the condition number of the local Hessian, and the rank of the data log-likelihood), works out from the physics of the field why the algorithm fails, and writes the code of a new operator; a candidate is kept only if it passes the check and gives uniform truth-quantile and likelihood ranks on held-out simulated events. Scientists specify the task, the compute budget, and the blinding rules (Supplemental Material). Three classes recur.

(i)~\textit{Physical coordinates and local chart.} Each algorithm works in coordinates $\bm{\xi}=\varphi(\bm{\theta})$ adapted to the physics---for orbits $h+ik=\sqrt{e}\,e^{i\omega}$, the mean longitude $\lambda_{\mathrm{ref}}=M_{\mathrm{ref}}+\omega$, and the orbital radius $r_{\mathrm{ref}}=a(1-e\cos E_{\mathrm{ref}})$ at a reference epoch, which remove the branch cuts at $e\to0$ and in the orbital phase and decouple $a$ from $e\cos E$; for X-ray spectra the logarithms of the component normalizations. Where the posterior is too concentrated for one global map, multistart optimization from prior draws gives the mode $\bm{\xi}_{\mathrm{MAP}}(\bm{x})$ and a Cholesky factor with $\mathbf{L}\mathbf{L}^\top=\mathbf{H}(\bm{x})^{-1}$. In the whitened coordinate
\begin{equation}
\bm{z}=\mathbf{L}(\bm{x})^{-1}\left[\varphi(\bm{\theta})-\bm{\xi}_{\mathrm{MAP}}(\bm{x})\right]
\label{eq:chart}
\end{equation}
the network $q_\psi(\bm{z}\mid\bm{x})$ is trained once, and the proposal density $q(\bm{\theta}\mid\bm{x})=q_\psi(\bm{z}\mid\bm{x})\,|\det\mathbf{L}|^{-1}\,|\det\partial\varphi/\partial\bm{\theta}|$ absorbs the volume contraction analytically, since $|\det\mathbf{L}|^{-1}=\sqrt{\det\mathbf{H}}\sim e^{D_{\mathrm{KL}}}$.

(ii)~\textit{Factorized importance sampling.} For $N$ stars that share an equation of state $\bm{\Lambda}$, the proposal is split into $q(\bm{\Lambda}\mid\bm{s}_{1:N})$ and per-star conditionals $q(\bm{\nu}_i\mid\bm{\Lambda},\bm{s}_i)$, with spectra $\bm{s}_i$ and stellar parameters $\bm{\nu}_i$, and, as in hierarchical inference~\cite{Hogg2010,Mandel2019,Steiner2018}, the weight of each equation-of-state draw is a product of per-star marginal estimates,
\begin{equation}
W_m\propto\frac{p(\bm{\Lambda}_m)}{q(\bm{\Lambda}_m\mid\bm{s}_{1:N})}\prod_{i=1}^{N}\frac{1}{K}\sum_{k=1}^{K}\frac{p(\bm{s}_i\mid\bm{\Lambda}_m,\bm{\nu}_{imk})\,p(\bm{\nu}_{imk})}{q(\bm{\nu}_{imk}\mid\bm{\Lambda}_m,\bm{s}_i)}.
\label{eq:factored_snis}
\end{equation}

(iii)~\textit{Response to missed mass.} When the check fails or heavy weight tails ($\hat{k}>0.7$) signal missed mass, the D1 algorithm draws fresh batches, and in D4 and D6 a few rounds of adaptive importance sampling refine every proposal.

The field-specific forms of these classes and the models, data, and parameters of each field are given in the Supplemental Material.

\end{document}